\documentclass[useAMS,usenatbib,usegraphicx]{mn2e}

\title[Constraining dark matter halo properties]{Constraining dark matter halo properties using lensed SNLS supernovae}
\author[J. J\"onsson et al.]{
J. J\"onsson$^{1}$\thanks{E-mail:jacke@astro.ox.ac.uk}, 
M. Sullivan$^{1}$,
I. Hook$^{1,2}$,
S. Basa$^{3}$, 
R. Carlberg$^{4}$, 
A. Conley$^{4}$,  \newauthor 
D. Fouchez$^{5}$,
D. A. Howell$^{6,7}$, 
K. Perrett$^{4}$, and 
C. Pritchet$^{8}$ \\ 
$^{1}$University of Oxford Astrophysics, Denys Wilkinson
  Building, Keble Road, Oxford OX1 3RH, UK \\
$^{2}$INAF -- Osservatorio Astronomico di Roma, via Frascati 33, 00040
Monteporzio (RM), Italy  \\
$^{3}$LAM,  Observatoire Astronomique de Marseille,
P\^ole de l'\'Eoile Site de Ch\^ateau--Gombert
38, rue Fr\'ed\'eric Joliot--Curie \\
13388 Marseille cedex 13, France\\
$^{4}$Department of Astronomy and Astrophysics, University of Toronto, Toronto, Ontario M5S 3H4, Canada \\
$^{5}$CPPM, CNRS--IN2P3 and Universit\'e Aix--Marseille II, Case 907, 13288 Marseille Cedex 9, France \\
$^{6}$Las Cumbres Observatory Global Telescope Network, 6740 Cortona Dr., Suite 102, Goleta, CA 93117, USA \\
$^{7}$Department of Physics, University of California, Santa Barbara, 
Broida Hall, Mail Code 9530, Santa Barbara, CA 93106--9530, USA \\
$^{8}$Department of Physics \& Astronomy, University of Victoria, Victoria, BC V8W 3P6, Canada 
 }
\begin{document}


\maketitle

\label{firstpage}

\begin{abstract}
This paper exploits the gravitational magnification of SNe~Ia to measure properties of 
dark matter haloes.
Gravitationally magnified and de-magnified SNe~Ia should be brighter and fainter than average, respectively. The magnification of individual SNe~Ia 
can be computed using observed properties of foreground galaxies and dark matter halo models. We model the dark matter haloes of the galaxies as truncated singular isothermal spheres with velocity dispersion and truncation radius obeying luminosity dependent scaling laws.

A homogeneously selected sample of 175 SNe~Ia from the first 3-years of the Supernova Legacy Survey (SNLS) in the redshift range $0.2 \la z \la 1$ is used to 
constrain models of the dark matter haloes associated with foreground galaxies.
The best-fitting velocity dispersion scaling law agrees well with galaxy--galaxy lensing measurements.
We further find that the normalisation of the velocity dispersion of passive and star forming galaxies are consistent with empirical  
Faber--Jackson and Tully--Fisher relations, respectively.
If we make no assumption on the normalisation of these relations, we find that the data prefer gravitational lensing at the 92 
per cent confidence level. 
Using recent models of dust extinction we deduce that the impact of this effect on our results is very small.

We also investigate the brightness scatter of SNe~Ia due to gravitational lensing, which has implications for SN~Ia cosmology. 
The gravitational lensing scatter is approximately proportional to the SN~Ia redshift. We find the constant of proportionality to be  $B \simeq 0.055_{-0.041}^{+ 0.039}$ mag
($B \la 0.12$ mag at the 95 per cent confidence level).
If this model is correct,
the contribution from lensing to the
intrinsic brightness scatter of SNe~Ia is small for the SNLS sample. 
According to the best-fitting dark matter model, gravitational lensing should, however, contribute significantly to the brightness scatter at $z \ga 1.6$. 
\end{abstract}


\section{Introduction} \label{sec:intro}
In the context of supernova cosmology, gravitational lensing is usually regarded as a source of uncertainty
because it adds extra scatter to the brightness of high redshift Type Ia supernovae  \citep{kan95,fri97,wam97,hol98,ber00,hol05,gun06}. 
Due to flux conservation, the effects of gravitational lensing magnification and de-magnification  average out  and is therefore expected to lead to negligible bias in cosmological parameter estimation 
\citep[see, e.g.,][]{sar08,jon08}.

Gravitationally lensed Type Ia supernovae (SNe~Ia) are, however,  interesting in their own right, because they can be used to measure magnification.
Magnified and de-magnified standard candles, like SNe~Ia calibrated using light curve shape and colour corrections, should appear to be brighter 
and fainter than average, respectively. 
Cosmic magnification has already been detected by \citet{scr05} using the correlation between quasars and galaxies observed within the Sloan Digital Sky Survey.
The magnification (or de-magnification) of SNe~Ia can be used to probe cosmology \citep{met99,dod06,zen09} and the nature of the 
lensing matter \citep{rau91,met98,metsil99,sel99,mor01,min02,met07,jon09}.

In addition to this paper there is another lensing study making use of Supernova Legacy Survey \citep[SNLS,][]{ast06} data.
The study of \citet{kro09}, which focus on the detection of a gravitational lensing signal using assumptions about the dark matter halo models, report the detection of  a signal at the 99 per cent confidence level. This result confirms an earlier tentative detection with GOODS data \citep{jon07}.
In the work presented here the assumptions of the underlying model are tested. 

We use presumably lensed high-redshift ($0.2 \la z \la 1$) SNe~Ia from the SNLS, 
to measure properties of dark matter
haloes of galaxies in the deep Canada--France--Hawaii Telescope Legacy Survey (CFHTLS) fields. The method we use was first suggested by \citet{met98}. Recently \citet{jon09} applied this method to SNe~Ia and galaxies observed within the Great Observatories Origins Deep Survey (GOODS). Since the SNLS SNe~Ia are not as distant ($z\la 1$) as the GOODS SNe~Ia, the effect of gravitational lensing is expected to be smaller than for the GOODS sample. However the SNLS SNe~Ia are far more numerous
and selected in a more homogeneous way.  

The data used to put constraints on dark matter halo properties are described in section~\ref{sec:data}. In section~\ref{sec:method} the method is 
described. The results are presented in section~\ref{sec:result}.  
We investigate the effect of dust extinction in section~\ref{sec:ext}.
Section~\ref{sec:impl} is devoted to 
gravitational lensing brightness scatter. 
The results are summarised  and discussed in section~\ref{sec:disc}. 

Throughout the paper we assume a cosmological model characterised by $\Omega_{\rm M}=0.27$, $\Omega_{\Lambda}=0.73$, and $h=0.7$. 

\section{Data} \label{sec:data} 
The SNLS is made up of photometric and spectroscopic observations. The
photometry is obtained as part of the deep component of CFHTLS with
the one square degree imager MegaCam \citep{bou03}. The deep part of
CFHTLS comprises four fields (D1, D2, D3, and D4), each
$\simeq1$\,deg$^2$ in size, imaged in $u^*$, $g^\prime$, $r^\prime$,
$i^\prime$, and $z^\prime$ filters approximately every 4--5 days during
dark and grey time, suitable for detecting supernovae and building
light curves \citep{ast06,guy09}. The spectroscopic observations are used to determine the
nature of supernova candidates and measure their redshifts \citep{how05,bro08,bal09}. To measure
properties of dark matter galaxy haloes using lensed SNe~Ia,
observations of both supernovae and galaxies are required.

\subsection{Galaxies}\label{sec:gal}
 The number of galaxies in each field is of the order $10^5$ and we
therefore have to rely upon photometric redshifts.
The techniques used 
to obtain galaxy properties
are described in detail in \citet{sul06}.
Galaxy photometry is performed using SExtractor \citep{ber96}
following \citet{sul06}. We estimate the galaxy redshift, $z_{\rm gal}$, and its
physical properties using the photometric redshift code Z--PEG, based
on the P\'EGASE.2 galaxy spectral evolution code
\citep{fioc97,leborgne02}. The best-fitting spectral energy distribution (SED)
 is determined using a
$\chi^2$ minimisation between the observed fluxes, the corresponding
flux errors, and the synthetic photometry generated by integrating the
template SEDs through the SNLS effective filter responses.

We then use this best-fitting SED for the estimation of the various
properties which can be used to characterise the galaxy. We first
require an appropriate luminosity. The galaxies could, in principle,
be characterised by the luminosity in any rest frame wavelength band,
but to facilitate comparison with results in the literature we use $M_B$, the
absolute magnitude in the 
$B$-band.  Given the photometric redshift of a galaxy and the
best-fitting SED, we compute absolute magnitudes by integrating the
de-redshifted SED through the $B$-band filter. 

We also require a galaxy type $\tau$ as we expect the relations
between dark matter haloes and galaxies to be different for different
galaxy types. 
As in \citet{sul06}, we use the specific star-formation rate (sSFR), the
star-formation rate (in $M_{\sun}$ yr$^{-1}$) per unit stellar mass.
If $\log({\rm sSFR})<-12$, the galaxy is classified as passive,
otherwise it is classified as star forming.

According to \citet{ilb06}, reliable photometric redshifts can only be obtained for galaxies
     with $m_i < 24$ mag from the CFHTLS deep data.
We therefore include only galaxies brighter
than this magnitude limit.   
The fraction of passive galaxies ranges
from 5 per cent in D1 and D3 to 8 per cent in
D4. 
We have checked that including galaxies fainter than
$m_{i^\prime}=24$ mag, which have less secure redshifts, have negligible
impact on the results. 

\subsection{Supernovae}
We assume that the brightness of a SN~Ia traces its magnification. In
our analysis we use the Hubble diagram residual,  
\begin{equation}
\Delta m_{\rm SN}=\mu_B-\mu(\Omega_{\rm M},\Omega_\Lambda;z_{\rm SN}),
\end{equation}
and its uncertainty, $\sigma_{\Delta m_{\rm SN}}$, to characterise the
brightness of a SN~Ia. 
In the equation above, $\mu_B$ is a distance indicator and $\mu(\Omega_{\rm M},\Omega_\Lambda;z_{\rm SN})$
is the predicted value for a SN~Ia at redshift $z_{\rm SN}$ assuming a cosmological model with  $\Omega_{\rm M}=0.27$ and 
$\Omega_\Lambda=0.73$. 
The Hubble diagram residual measures the
brightness of an individual SN relative to the average of the
population -- brighter SNe have more negative residuals.

Our set of residuals was computed using the SiFTO \citep{con08} light curve fitting package.
We have also performed the
analysis with residuals computed using the SALT2 \citep{guy07} 
package. The results obtained using SiFTO and SALT2 residuals are very similar and  we therefore only present results obtained using SiFTO residuals here.
For this set of residuals the intrinsic scatter, i.e.~the component of the scatter which cannot be attributed to known observational errors, is approximately 0.09 mag \citep{guy09}.

In order to compute the magnification due to the galaxies in the
foreground we also need, in addition to the redshift, $z_{\rm SN}$,  
the position on the
sky, $\btheta_{\rm SN}$, for each SN~Ia. When available we use the 
spectroscopic redshift of the host galaxy, otherwise the spectroscopic 
redshift of the SN~Ia itself. Positions of the SNe~Ia, as well as of the 
galaxies, were obtained from the CFHTLS $i^\prime$-band photometry.

We use 243 SNe Ia from the 3-year SNLS data set \citep{guy09}.
Since some parts of the deep CFHTLS
fields are covered by bright stars, they have to be masked \citep{sul06}. SNe~Ia
located too close to a masked region or the boundary of the field are
removed from the sample because of the lack of observations of
foreground galaxies. 
To characterise the size of a masked
region we use an effective radius, $\theta_{\rm eff}=\sqrt{A/\pi}$,
where $A$ is the area of the masked region.  The effective radius
ranges from a few to a few hundred arc seconds. 
Roughly
20 per cent of the deep fields are covered by masked regions, but most of this
area is made up of a few regions with $\theta_{\rm eff}>60\arcsec$.

When computing the magnification, all galaxies within a radius $\theta_{\rm c}$ are included. 
We use $\theta_{\rm c}=60\arcsec$, which according to simulations \citep{gun06,kro09}, is sufficient. 
For a SN~Ia to be included in the analysis we therefore
use the selection criterion that the distance to the boundary of the field or the distance to the boundary of the nearest masked region must be larger than $\theta_{\rm c}$. If all masked regions were taken into account, most SNe~Ia would lie too close to a masked region and for that reason fail the selection criterion. We therefore ignore all masked regions with $\theta_{\rm eff}<20\arcsec$. 
Since the location of the masked regions should be uncorrelated with the location of the foreground galaxies, ignoring small masked regions should
not bias the results, only increase the uncertainty.

Only 175 out of the 243 SNe~Ia fulfill the selection criterion. 
In Table~\ref{tab:nsn} the total number of SNe~Ia together 
with the number of selected ones
 are  listed for each field. 
 
\begin{table}
\caption{Number of SNe~Ia 
in the CFHTLS deep fields.}              
\label{tab:nsn}      
\centering                                      
\begin{tabular}{c c c c}          
\hline\hline                        
Field & All SNe~Ia & Selected SNe~Ia\\   
\hline                                   
    D1 &  55 & 42 \\      
    D2 & 49 & 33 \\
    D3 & 75 & 55 \\
    D4 & 64 & 45 \\
    Total & 243 & 175 \\
\hline                                             
\end{tabular}
\end{table}

\section{Method} \label{sec:method}
The method \citep{met98} we use can be summarised in two steps. First the magnifications of the SNe~Ia are computed for a range of halo models.
Then the best-fitting halo model to the Hubble diagram residuals is obtained by $\chi^2$ minimisation.
In the following the method is described in more detail.

\subsection{Computing the magnification}
In order to constrain properties of dark matter haloes we exploit the correlation between the brightness of SNe~Ia and their magnification.
For each SN~Ia the gravitational magnification due to the dark matter haloes associated with the 
observed galaxies in the foreground, $\Delta m_{\rm lens}(\blambda_{\rm SN},\blambda_{\rm los};\blambda_{\rm halo})$, 
is computed assuming a halo model with parameters described by $\blambda_{\rm halo}$. 
To describe the redshift, $z_{\rm SN}$, and location, $\btheta_{\rm SN}$, of the SN~Ia, which are used in the calculation, we use the shorthand notation 
\begin{equation}
\blambda_{\rm SN}=\{z_{\rm SN},\btheta_{\rm SN}\}.
\end{equation}
For the calculation of the magnification we also need properties of the $N_{\rm gal}$ galaxies along the SN~Ia 
line of sight, which we denote by
\begin{equation}
\blambda_{\rm los}=\{z_{\rm gal}^1,\btheta_{\rm gal}^1,M_B^1,\tau^1,\ldots, 
z_{\rm gal}^{N_{\rm gal}},\btheta_{\rm gal}^{N_{\rm gal}},M_B^{N_{\rm gal}},\tau^{N_{\rm gal}} \},
\end{equation}
where $z_{\rm gal}^i$, $\btheta_{\rm gal}^i$, $M_B^i$, and $\tau^i$ are redshift, position on the sky, absolute $B$-band magnitude, and type of the $i$th galaxy.

To compute the magnification 
of a SN~Ia we use
the weak lensing approximation \citep[see, for example,][]{sch92}. 
We have checked the validity of this approximation against a more time consuming ray tracing algorithm for a few cases. 
For the sample of SNe~Ia considered here, which have relatively low redshifts, the difference is found to be $\la 5$ per cent.

According to the weak lensing approximation the magnification, in terms of magnitudes, is related to the 
convergence, $\kappa$, (which is a dimensionless surface density) via
\begin{equation}
\Delta m_{\rm lens}\simeq -2.17\kappa,
\end{equation}
where 
\begin{equation}
\kappa=\kappa_{\rm los}-\kappa_{\rm b}.
\label{eq:kapsum}
\end{equation}
The first term in equation~(\ref{eq:kapsum}) is a sum over the contribution to the convergence from each galaxy, $\kappa_{\rm gal}$, along the line of sight,
\begin{equation}
\kappa_{\rm los}=\sum_{i=1}^{N_{\rm gal}} \kappa_{\rm gal}^i.
\label{eq:kapgal}
\end{equation}
The second term, representing the compensating effect of the background density, acts as a normalisation allowing the magnification relative to a homogeneous universe to be computed. Due to flux conservation $\langle \kappa \rangle=0$. To ensure that this condition is fulfilled we use 
$\kappa_{\rm b}=\langle \kappa_{\rm los} \rangle$. The average value of $\kappa_{\rm los}$ is computed for a large number of randomly 
selected lines of sight. 
One important element of the model remains to specify, namely $\kappa_{\rm gal}$.  

\subsection{Halo models} \label{sec:halo}
The contribution to the dimensionless convergence from each galaxy is given by 
\begin{equation}
\kappa_{\rm gal}(\bxi)=\frac{\Sigma(\bxi)}{\Sigma_{\rm c}}, 
\label{eq:kapdef}
\end{equation}
where the surface density, $\Sigma(\bxi)$, is obtained by projecting the matter distribution onto a lens plane,
\begin{equation}
\Sigma(\bxi)=\int_{-\infty}^{\infty} \rho(\bxi,y)dy,
\label{eq:proj}
\end{equation} 
where $\bxi$ is a vector in the plane and $y$ is a coordinate along the line of sight. This is the step
in the calculations where the density profile, $\rho(\mathbf{r})$, of the dark matter halo enters. 
Equation~(\ref{eq:kapdef}) depends also on the critical surface density,
\begin{equation}
\Sigma_{\rm c}=\frac{1}{4\pi G}\frac{D_{\rm s}}{D_{\rm d}D_{\rm ds}},
\label{eq:scrit}
\end{equation}
which, in turn, depends on the angular diameter distances between the observer and the source, $D_{\rm s}$,
the observer and the lens (deflector), $D_{\rm d}$, and the lens and the source, $D_{\rm ds}$. 
The distances are computed from $z_{\rm SN}$ and $z_{\rm gal}$ assuming a cosmological model 
($\Omega_{\rm M}=0.27$, $\Omega_\Lambda=0.73$, and $h=0.7$). 

We assume that the dark matter haloes can be described by a singular isothermal sphere (SIS). The density profile of a SIS,
\begin{equation}
\rho(r)=\frac{\sigma^2}{2\pi G}\frac{1}{r^2},
\end{equation}
has  only one free parameter -- the velocity dispersion, $\sigma$.
The formula for the convergence of a SIS can be considerably simplified, if we introduce a suitable length scale $\xi_0$ in the lens plane,
\begin{equation}
\xi_0=4\pi  \sigma^2 \frac{D_{\rm d}D_{\rm ds}}{D_{\rm s}}.
\label{eq:sisxi}
\end{equation}
Since the SIS profile is divergent, the profile is truncated at a radius $r_{\rm t}$. The convergence of a truncated SIS is 
given by
\begin{equation}
\kappa_{\rm gal}(x)=\left\{ 
\begin{array}{ll}
\frac{1}{\pi x}\arctan\sqrt{ {x_{\rm t}^2}/{x^2}-1} &\mbox{ if $x \leq x_{\rm t}$} \\
 0  &\mbox{ if $x > x_{\rm t}$},
       \end{array}
\right.
\label{eq:kapsis}
\end{equation}
where $x=\xi/\xi_0$ and $x_{\rm t}=r_{\rm t}/\xi_0$.

To investigate the relationship
between galaxy luminosity and velocity dispersion, we use the scaling law
\begin{equation}
\sigma=\sigma_{*}\left( \frac{L}{L_*} \right)^{\eta},
\label{eq:sislaw}
\end{equation}
where $L_*$ is a fiducial luminosity, which we take to be $L_*=10^{10}h^{-2}L_{\sun}$ in the $B$-band. 
In terms of absolute $B$-band magnitudes, which we will work with, the scaling relation becomes
\begin{equation}
\sigma=\sigma_*10^{-\eta(M_B-M_B^*)/2.5}, 
\end{equation}
where $M_B^*=-19.52+5\log_{10} h$. 

For the truncation radius we will consider a scaling law  of the form
\begin{equation}
r_{\rm t}=r_*\left(\frac{\sigma}{\sigma_*}\right)^{\gamma}=r_*\left(\frac{L}{L_*}\right)^{\eta\gamma}.
\label{eq:trlaw}
\end{equation}
Since we only include galaxies located a distance $\theta_{\rm c}$ from the position of the SN~Ia, the truncation radius has an effect only if
$r_{\rm t}/D_{\rm d}<\theta_{\rm c}$. For $\theta_{\rm c}=60\arcsec$ and $z_{\rm gal}<1$ this corresponds to $r_{\rm t} \la 300h^{-1}$ 
kpc.

\subsection{Finding the best-fitting model}
In summary, the parameters of our halo model are
\begin{equation}
\blambda_{\rm halo}=\{\sigma_*,\eta,r_*,\gamma\},
\end{equation}
where $\{ \sigma_*,\eta\}$ and  $\{r_*,\gamma\}$ describe the velocity dispersion and truncation radius scaling law, respectively. 
To find the best-fitting halo model we minimise the following $\chi^2$-function
\begin{equation}
\chi^2=\sum_{i=1}^{N_{\rm SN}}
\frac{\left[ \Delta m_{\rm SN}^i-\Delta m_{\rm lens}(\blambda_{\rm SN}^i,\blambda_{\rm los}^i;
\blambda_{\rm halo})\right]^2}
{\sigma_{\rm int}^2+(\sigma_{\Delta m_{\rm SN}}^i)^2},
\label{eq:chi2}
\end{equation}
where the superscript $i$ refers to the $i$th SN~Ia.

In equation~(\ref{eq:chi2}) the intrinsic dispersion, $\sigma_{\rm int}$, is added in quadrature to the uncertainty of the Hubble diagram residual. 
The uncertainty in $\Delta m_{\rm lens}(\blambda_{\rm SN},\blambda_{\rm los};\blambda_{\rm halo})$,
which for a specific halo model $\blambda_{\rm halo}$ only depends on the uncertainties in $\blambda_{\rm SN}$ and $\blambda_{\rm los}$,
 is negligible compared to
$\sigma_{\rm int}$ and $\sigma_{\Delta m_{\rm SN}}$ and is for that reason neglected in equation~(\ref{eq:chi2}).

\subsection{Cosmic variance}
\begin{figure} 
\includegraphics[width=84mm,angle=-90]{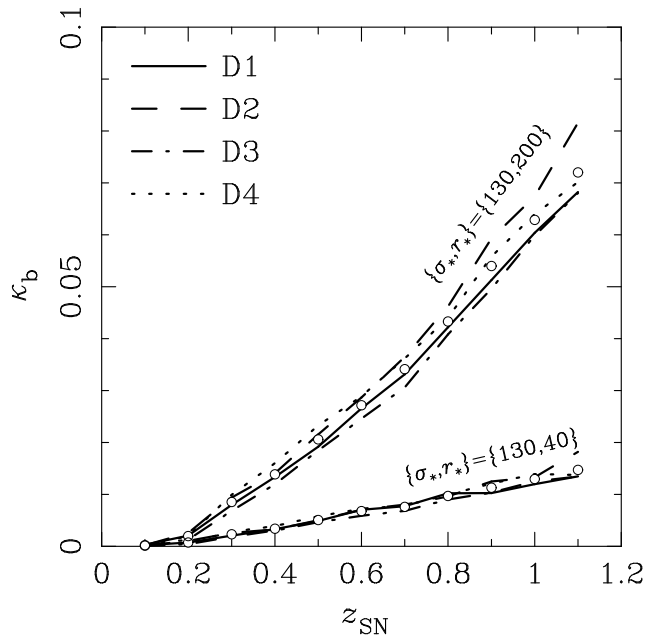}
\caption{
Convergence normalisation, $\kappa_{\rm b}$, as a function of source redshift. Solid, dashed, dash-dotted, and dotted curves correspond to D1, D2, D3, and D4, respectively. The two sets of curves are computed for two different SIS models. For both models  $\eta=1/3$ and $\gamma=0$,
but the values of  $\sigma_*$ (measured in km s$^{-1}$) and $r_*$ (measured in $h^{-1}$ kpc) differ. 
 The lower and upper set of curves correspond to $\{\sigma_*,r_*\}=\{130,200\}$ and 
 $\{\sigma_*,r_*\}=\{130,40\}$, respectively. Open circles indicate the average of $\kappa_{\rm b}$ computed for the four fields.
}
\label{fig:norm} 
\end{figure}

The four deep CFHTLS  fields are located widely separated in the sky. 
Due to cosmic variance we expect the density of galaxies to vary between the fields. 
This variation affects the normalisation of the convergence, $\kappa_{\rm b}$. 
Figure~\ref{fig:norm} shows $\kappa_{\rm b}$ as a function of source redshift for the four fields computed for two different SIS models.
For both models $\eta=1/3$ and $\gamma=0$.
The lower set of curves correspond to $\sigma_*=130$ km s$^{-1}$ and $r_*=40h^{-1}$ kpc, while the upper set of curves
corresponds to $\sigma_*=130$ km s$^{-1}$ and $r_*=200h^{-1}$ kpc.
For both models the differences between the fields increase with source redshift.
In order to reduce the effect of cosmic variance, we use the average value of $\kappa_b$, as indicated by the circles

\section{Results} \label{sec:result}
The halo model, $\blambda_{\rm halo}$, has four parameters. Instead of constraining all parameters simultaneously, which would be quite time consuming, we focus 
on different aspects of the model in the following.

\subsection{Truncation radius}
Our model of the luminosity dependent truncation radius, equation~(\ref{eq:trlaw}), has two parameters:
$r_*$ and $\gamma$. 
In order to reduce the size of the parameter space to a manageable size, the investigation is limited to a single value of $\eta$ in equation~(\ref{eq:sislaw}).
We expect this parameter to be in the range
$1/4 \la \eta \la 1/3$, since $\eta \simeq 1/4$ and $\eta \simeq 1/3$ are the exponents belonging to the 
Faber--Jackson \citep[F--J,][]{fab76} and Tully--Fisher \citep[T--F,][]{tul77} relations valid for early and late type galaxies, respectively. Since most of the galaxies in our sample are star forming and therefore should be of late type, we assume $\eta=1/3$.
We have checked that the dependence of $\eta$ is rather weak, justifying our restricted investigation.

When minimising equation~(\ref{eq:chi2}) with respect to the three parameters we find two minima with similar $\chi^2$. For the first minimum 
($\sigma=114$ km s$^{-1}$, $r_*=170h^{-1}$ kpc, $\gamma=-2.8$) the value of the $\chi^2$ is 
164.50. The $\chi^2$ for the second minimum ($\sigma=124$ km s$^{-1}$, $r_*=69h^{-1}$ kpc, $\gamma=-0.9$) is  slightly higher, 164.66. 
The best-fitting velocity dispersion normalisations, $\sigma_*$, are similar for the two minima. 
Negative values of $\gamma$ correspond to rather peculiar truncation radius scaling laws. For these scaling laws more luminous, and presumably more massive, galaxies have smaller truncation radii than less luminous galaxies.
Unfortunately, the data do not allow us to put any limits on $r_*$ and $\gamma$. 

\begin{figure} 
\includegraphics[width=84mm,angle=-90]{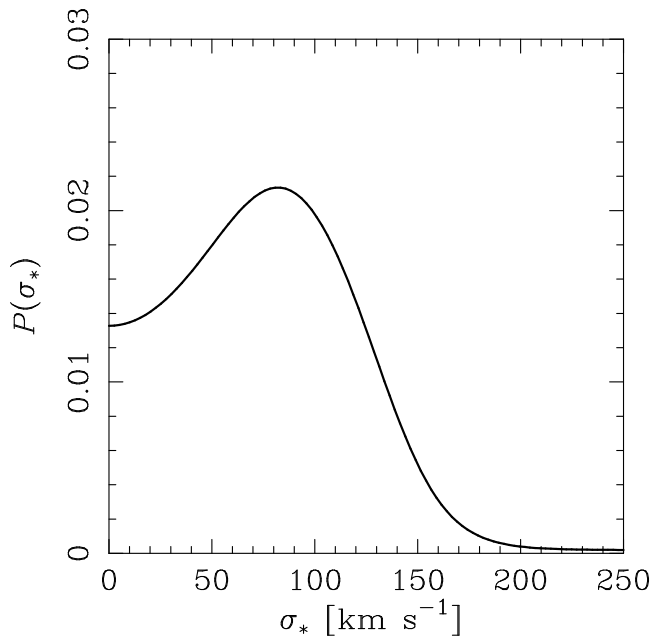}
\caption{Probability density function of $\sigma_*$ after marginalisation over 
$r_*$ and $\gamma$. 
The exponent of the luminosity dependent 
velocity dispersion scaling law is $\eta=1/3$.
}
\label{fig:sig} 
\end{figure}

Fortunately, limits can be obtained for the normalisation of the velocity dispersion. 
Figure~\ref{fig:sig} shows the 
probability density function of $\sigma_*$ after marginalisation over $r_*$ and $\gamma$. 
We find $\sigma_*=83^{+37}_{-56}$ km s$^{-1}$ 
(68.3 per cent confidence level) and $\sigma_* < 147$ km s$^{-1}$ at 95 per cent confidence level. 

Let us now compare the best-fitting values of the truncation radius 
with the virial radius.
The virial radius is defined as the radius inside which the mean density of the halo is $\Delta_{\rm vir}(z)$ times the mean matter density, 
$\rho_{\rm m}(z)=\Omega_{\rm M}\rho_{\rm c}^0(1+z)^3$, where $\rho_{\rm c}^0$ is the present critical density. 
To compute the virial over density we use the approximation of \citet{bry98},
\begin{equation} 
\Delta_{\rm vir}(z) \simeq (18\pi^2+82x-39x^2)\rho_{\rm c}(z)/\rho_{\rm m}(z), 
\end{equation}
where $x=\rho_{\rm m}(z)/\rho_{\rm c}(z)-1$.
For a SIS profile the virial radius is
\begin{equation}
r_{\rm vir}=20\sigma \left[ \Delta_{\rm vir}(z)\Omega_{\rm M}(1+z)^3 \right]^{-1/2} h^{-1} \mbox{\,kpc}.
\label{eq:rvir}
\end{equation}
From equations~(\ref{eq:kapdef}) and (\ref{eq:scrit}) it is clear that there is an optimum lens redshift, $z_{\rm gal}$, for every source 
redshift, $z_{\rm SN}$,
depending on the distance combination $D_{\rm d}D_{\rm ds}/D_{\rm s}$, where the convergence is maximal. Our distribution of sources peaks at 
$z_{\rm SN} \simeq 0.7$, which corresponds to an optimum lens redshift of $z_{\rm gal} \simeq 0.3$.  For a dark matter halo at this redshift, the virial 
radius is, according to equation~(\ref{eq:rvir}),  $r_{\rm vir} \simeq 1.6\sigma h^{-1}$ kpc. 

The virial radius of an $L_*$ galaxy, for which we find $\sigma_* \simeq 120$ km s$^{-1}$, is hence
$r^*_{\rm vir} \simeq 190h^{-1}$ kpc. 
When exploring the truncation radius scaling law we find $r_*=170h^{-1}$ kpc, which is similar to the virial radius, but the exponent is peculiar, 
$\gamma=-2.8$. 
As we will see in the next section, we find even smaller values of $r_*$
for $\gamma=0$.

\subsection{Velocity dispersion scaling law}
\begin{figure} 
\includegraphics[width=84mm,angle=-90]{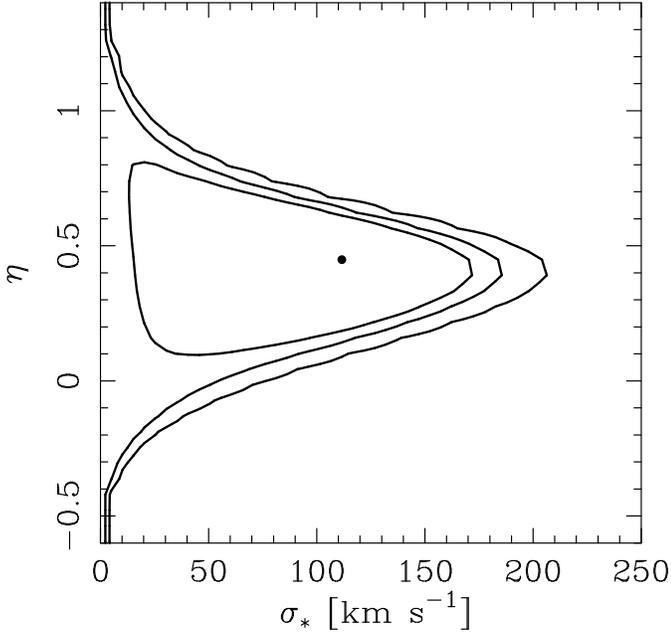} 
\caption{Best-fitting values (circle) and 68.3, 95, and 99 per cent confidence level contours for the luminosity dependent  velocity dispersion 
scaling law after marginalisation over $r_*$. 
}
\label{fig:fit1} 
\end{figure}

Let us now investigate the velocity dispersion scaling law, which is described by the parameters $\sigma_*$ and $\eta$. 
Here we consider a constant truncation radius, $r_{\rm t}=r_*$, i.e.~we assume $\gamma=0$. 
The best-fitting values of the three parameters to the data are $\sigma_*=122$ km s$^{-1}$, $\eta=0.45$, and $r_*=40h^{-1}$ kpc.  
This value of $r_*$ is only one fifth of the virial radius corresponding to $\sigma_*$. 
For this minimum the $\chi^2$ is 164.30.
In order to obtain constraints on the velocity dispersion scaling law ($\sigma_*$ and $\eta$), we marginalise over $r_*$. Reasonable fits to the data can not be obtained for all values of $r_*$ and we therefore apply a flat prior, 
\begin{equation}
\pi(r_*)=\left\{ 
\begin{array}{ll}
 1&\mbox{ if $20h^{-1} < r_* <100h^{-1}$} \\
 0  &\mbox{ otherwise},
       \end{array}
\right.
\end{equation}
 when marginalising over this parameter. 
Figure~\ref{fig:fit1} shows the results for the velocity dispersion 
scaling law. 
The circle indicates the best-fitting values ($\sigma_*=112$ km s$^{-1}$ and  $\eta=0.45$).
In the figure are also shown confidence level contours at the 68.3, 95, and 99 
per cent confidence level. 

\begin{figure} 
\includegraphics[width=84mm,angle=-90]{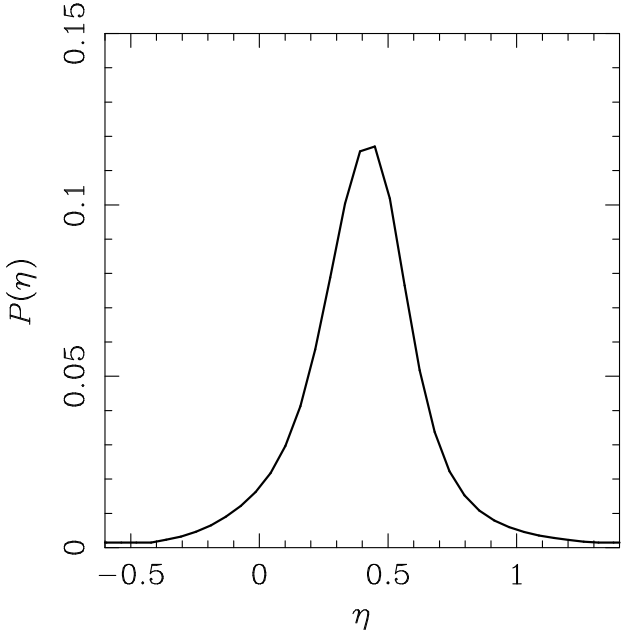}
\caption{Probability density function of  $\eta$ after marginalisation over $\sigma_*$ and $r_*$.
}
\label{fig:eta} 
\end{figure}

Figure~\ref{fig:eta} shows the probability density function of $\eta$ after marginalisation over $\sigma_*$ and $r_*$. We find 
$\eta=0.45^{+0.19}_{-0.27}$ (68.3 per cent confidence level) and $-0.34 < \eta < 1.17$ at the 95 per cent confidence level. 
For the velocity dispersion normalisation we can only obtain upper limits. After marginalisation over $\eta$ and $r_*$ 
we find $\sigma_* < 156$ at the 95 per cent confidence level.

We have redone the fit of the velocity dispersion scaling law assuming the truncation radius to 
be equal to the virial radius, given by equation~(\ref{eq:rvir}). 
The best-fitting model ($\chi^2=168.13$) is in this case $\sigma_*=56$ km s$^{-1}$
and $\eta=0.25$. 
 After marginalisation we find $\sigma_* < 91$ km s$^{-1}$ at the 95 per cent confidence level
 and $\eta=0.25^{+0.24}_{-0.29}$ ($-0.27 < \eta <1.24$ at the 95 per cent confidence level).

\subsection{Disentangling the velocity dispersion of different galaxy types} \label{sec:dis}
As already noted we expect different scaling laws for different galaxy types. The galaxies in our sample are classified as either passive or star forming 
depending on their sSFR (see section~\ref{sec:gal}). 
We will associate passive and star forming galaxies with early and late type galaxies, respectively. 
For passive and star forming galaxies we assume the velocity dispersion to scale with luminosity as 
\begin{equation}
\sigma^{\rm p}=\sigma^{\rm p}_*\left(\frac{L}{L_*}\right)^{1/4}
\end{equation}
and
\begin{equation}
\sigma^{\rm sf}=\sigma^{\rm sf}_*\left(\frac{L}{L_*}\right)^{1/3},
\end{equation}
respectively.
The exponents of these scaling laws correspond to the F--J and T--F relations.
Once again we restrict our study to models with $\gamma=0$. 

\begin{figure} 
\includegraphics[width=84mm,angle=-90]{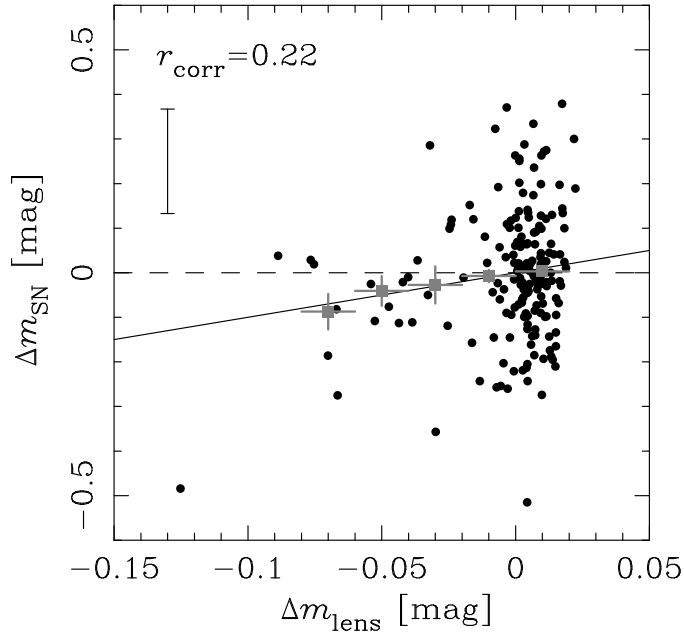} 
\caption{
Magnification-residual diagram for the best-fitting halo model with different scaling laws for passive and star forming galaxies. 
Solid squares represent the weighted average residual computed for SNe~Ia in magnification bins of width 0.02 mag.
An average residual error bar is shown in the upper left corner. The slope of the solid line is unity.
}
\label{fig:corr} 
\end{figure}

For the best-fitting halo model, $\sigma_*^{\rm p}=196$ km s$^{-1}$, 
$\sigma_*^{\rm sf}=109$ km s$^{-1}$,  and $r_*=45h^{-1}$ kpc, the $\chi^2$ is 164.59.
Figure~\ref{fig:corr} shows a 
magnification-residual diagram computed for this  model. 
We expect the points to scatter around the solid line with slope unity. The scatter around this line is 0.147 mag and
the linear correlation coefficient is  $r_{\rm corr}=0.22$. 
For the Hubble diagram residuals the scatter is  0.150 mag. Correcting  this sample of SNe~Ia for lensing \citep{gun06,jon09b} using the best-fitting model would hence lead to a very small reduction of the scatter.

\begin{figure} 
\includegraphics[width=84mm,angle=-90]{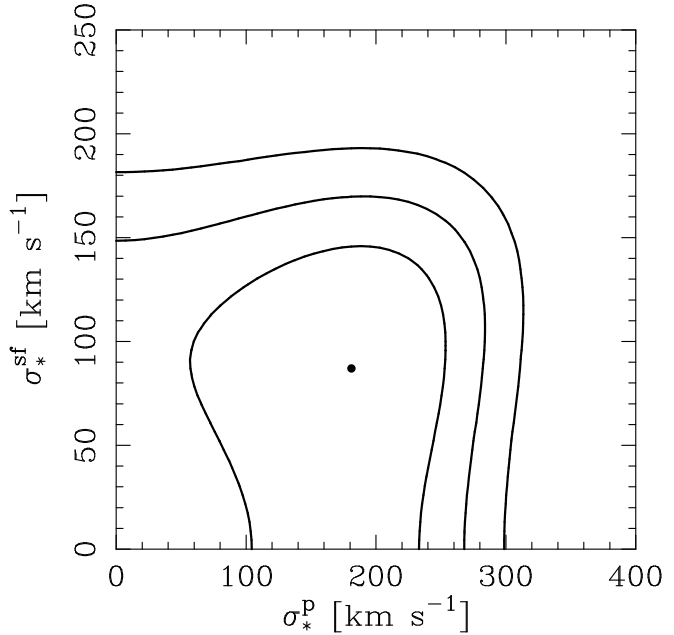} 
\caption{Best-fitting values (circle) and 68.3, 95, and 99 per cent confidence level contours for the velocity dispersion normalization for passive 
($\sigma_*^{\rm p}$)
and star forming ($\sigma_*^{\rm sf}$) galaxies after marginalisation over $r_*$. 
}
\label{fig:fit2} 
\end{figure}

We marginalise over  $r_*$ using the prior $\pi(r_*)$ to find constraints on $\sigma^{\rm p}_*$ and $\sigma^{\rm sf}_*$. 
Figure~\ref{fig:fit2} shows the result of our effort to disentangle the velocity dispersion for different galaxy types. The best-fitting values ($\sigma_*^{\rm p}=181$ and  $\sigma_*^{\rm sf}= 87$ km s$^{-1}$) after marginalisation over $r_*$ are indicated by the circle. 
The point $\sigma_*^{\rm p}=\sigma_*^{\rm sf}=0$, corresponding to no lensing, is excluded at the 91.6 per cent confidence level. 
This detection is in good agreement with the results of \citet{jon08}, where the probability of detecting  lensing at this confidence level for a sample of 175 SNLS SNe~Ia was predicted to be 80--90 per cent.
 
From Fig.~\ref{fig:fit2} it is clear that larger values of the velocity dispersion are admitted for passive than for star forming galaxies. 

If we marginalise over $r_*$ and  $\sigma_*^{\rm sf}$  ($\sigma_*^{\rm p}$), we can obtain limits on $\sigma_*^{\rm p}$ ($\sigma_*^{\rm sf}$).
We find  $\sigma_*^{\rm p}=181^{+60}_{-86}$ 
($\sigma_*^{\rm sf}=91^{+36}_{-63}$) km s$^{-1}$ (68.3 per cent confidence level) and 
$\sigma_*^{\rm p} < 260$ ($\sigma_*^{\rm sf} < 155$) km s$^{-1}$ at the 95 per cent confidence level.

When assuming $r_{\rm t}=r_{\rm vir}$ we find the
best-fitting velocity dispersion normalisation for passive and star forming galaxies ($\chi^2=167.49$) to be $\sigma_*^{\rm p}=113$ and 
 $\sigma_*^{\rm sf}=33$ km s$^{-1}$, respectively. After marginalisation we find
  $\sigma_*^{\rm p}=109^{+42}_{-70}$ km s$^{-1}$ 
($\sigma_*^{\rm p} < 172$ km s$^{-1}$ at the 95 per cent confidence level) for passive galaxies. For star forming galaxies we obtain an upper limit
at the 95 per cent confidence level after marginalisation,
$\sigma_*^{\rm sf} < 101$ km s$^{-1}$. 
Evidently, the increased truncation radius results in lower values  of the velocity dispersion normalisations. At the 99 per cent confidence level we find 
$\sigma_*^{\rm p}<195$ and $\sigma_*^{\rm sf} < 117$ km s$^{-1}$.

The results obtained in this section are summarised in Tables~\ref{tab:sum} and~\ref{tab:sum2}.

\begin{table*}
\caption{Summary  of best-fitting models.} 
\label{tab:sum}      
\centering                                      
\begin{tabular}{l c c c c c}          
\hline\hline                        
Studied aspect of the model&\multicolumn{2}{c}{Velocity dispersion scaling law} & \multicolumn{2}{c}{Truncation radius scaling law} \\ 
 & $\eta$ & $\sigma_*$ (km s$^{-1}$)& $\gamma$ & $r_*$ ($h^{-1}$ kpc)& $\chi^2$ \\
\hline                                   
No lensing & \multicolumn{2}{c}{$\sigma=0$}  & & & 168.66 \\
 Truncation radius scaling law (first minimum)&  $0.33$ (fixed) & 114 & $-2.8$ & 170 & 164.50 \\      
  Truncation radius scaling law (second minimum) &  $0.33$ (fixed) & 124 & $-0.9$ & 69 & 164.66 \\    
  Velocity dispersion scaling law &  $0.45$  & 122 & 0 (fixed) & 40 & 164.30 \\    
   Velocity dispersion scaling law &  $0.25$  & 56 & 
    \multicolumn{2}{c}{$r_{\rm t}=r_{\rm vir}$} & 168.13 \\    
Star forming/passive velocity dispersion scaling law & $0.33/0.25$ (fixed)& $109/196$ & 0 (fixed) & 45 &164.59 \\
Star forming/passive velocity dispersion scaling law & $0.33/0.25$ (fixed)& $33/113$ & \multicolumn{2}{c}{$r_{\rm t}=r_{\rm vir}$} & 167.49 \\
\hline                                             
\end{tabular}
\end{table*}

\begin{table*}
\caption{Summary  of constrains (with confidence level within parenthesis) on velocity dispersion scaling laws after marginalisation.}
\label{tab:sum2}      
\centering                                      
\begin{tabular}{l c c}          
\hline\hline                        
Studied aspect of the model&\multicolumn{2}{c}{Velocity dispersion scaling law}  \\ 
 & $\eta$ & $\sigma_*$ (km s$^{-1}$) \\
\hline                                   
 Truncation radius scaling law &  $0.33$ (fixed)& $83^{+37}_{-56}$ (68.3 p.c.) \\      
  Velocity dispersion scaling law ($\gamma=0$)&  $0.45^{+0.19}_{-0.27}$ (68.3 p.c.)  & $<156$ (95 p.c.)  \\    
   Velocity dispersion scaling law ($r_{\rm t}=r_{\rm vir}$) &  $0.25^{+0.24}_{-0.29}$ (68.3 p.c.) & $<91$ (95 p.c.) \\
Star forming velocity dispersion scaling law ($\gamma=0$) &  $0.33$ (fixed)& $91^{+36}_{-63}$ (68.3 p.c.)  \\
Passive velocity dispersion scaling law ($\gamma=0$) &  $0.25$ (fixed)& $181^{+60}_{-86}$ (68.3 p.c.)  \\
Star forming velocity dispersion scaling law ($r_{\rm t}=r_{\rm vir}$) & $0.33$ (fixed)& $<101$ ($95$ p.c.)  \\
Passive velocity dispersion scaling law ($r_{\rm t}=r_{\rm vir}$) & $0.25$ (fixed)& $109^{+42}_{-70}$ ($68.3$ p.c.)  \\
\hline                                             
\end{tabular}
\end{table*}

\section{Extinction by dust} \label{sec:ext}
The brightness of SNe~Ia can be affected by dust extinction in addition to gravitational lensing. Extinction by dust, in contrast to gravitational lensing, leads only to dimming of SNe~Ia. Extinction could therefore lead to underestimation of the effect of gravitational lensing magnification. 

Recently \citet{men09a} used the correlation between the brightness of tens of thousands of quasars and the position of tens of millions of galaxies in the foreground to measure magnification and extinction simultaneously. According to their measurements the average extinction and magnification have the same dependence on the impact parameter ($\propto \xi^{-0.8}$). 

Dust extinction and gravitational lensing can be separated because the former effect is wavelength dependent while the latter is achromatic.
In order to calibrate the SNe~Ia to become standard candles, the $B$-band magnitude, $m_B$, is corrected \citep{ast06},
\begin{equation}
\mu_B=m_B-\mathcal{M}+\alpha(s_B-1)-\beta\mathcal{C},
\label{eq:sifto}
\end{equation}
using stretch, $s_B$, which parameterizes light curve shape, and colour, $\mathcal{C}$.
The parameters $\mathcal{M}$, $\alpha$, and $\beta$ are determined from the data.
The colour correction could potentially correct for some of the dust extinction. 
According to \citet{men09b}, who have investigated the consequences of dust extinction for SN~Ia 
cosmology, this colour correction is not sufficient.

The values of the truncation radius preferred by the data are smaller than the virial radius 
(see section~\ref{sec:result}).
This rather surprising result might be explained by dust extinction not accounted for by the colour correction in equation~(\ref{eq:sifto}). 
Another indication of the small values of $r_{\rm t}$ preferred by the data, is the small values of $\sigma_*^{\rm p}$ and $\sigma_*^{\rm sf}$ obtained when assuming $r_{\rm t}=r_{\rm vir}$ (see the end of section~\ref{sec:dis}).  We have investigated if extinction by dust could explain the small values
of these halo parameters.

First we use the high $A_B$ model  in \citet{men09b}, which only depends on $z_{\rm SN}$ (see their fig.~1),
to  estimate the extinction in the $B$-band, $A_B$, for each SNe~Ia. The values of $A_B$ are then used to correct $m_B$ and $\mathcal{C}$ [this variable is essentially $E(B-V)$] assuming a standard dust law. 
For the high $A_B$ model $R_B=R_V+1=4.9$.
Thereafter we fit 
$\Omega_{\rm M}$, $\mathcal{M}$, $\alpha$, and $\beta$ to the extinction corrected data and compute new extinction corrected residuals. 
These parameters are hence allowed to compensate for the effect of dust extinction.
A flat Universe is assumed, i.e.~$\Omega_\Lambda=1-\Omega_{\rm M}$.
In table~\ref{tab:fit} the differences in these parameters with respect to the uncorrected case are listed.
Correcting the SNe~Ia for redshift dependent extinction has clearly a very small effect on the parameters, and consequently on the residuals.
\citet{men09b} also find small differences in $\mathcal{M}$, $\alpha$, and $\beta$ (see their table 1). For $\Omega_{\rm M}$, however, we find only half of their difference ($\Delta \Omega_{\rm M}=0.017$).
The statistical significance of the difference is, however, very similar: they find $0.55\sigma$, whereas we find $0.50\sigma$.  
Note that the parameter $\beta \simeq 3$ \citep{tri98} is different from $R_B=4.1$ measured for Milky Way dust \citep{sav79,sea79,rie85,car89}. 

Finally, we redo the fit of the halo parameters $\sigma_*^{\rm p}$ and $\sigma_*^{\rm sf}$ using the extinction corrected residuals. 
We find the difference in the best-fitting values of the halo parameters due to the extinction correction to be negligible. 
This is not very surprising because the difference between the extinction corrected and uncorrected residuals is on average 
only $0.000 \pm 0.005$ mag.

Since we have information about the foreground galaxies, the effect of a more detailed extinction correction, where the contribution from each individual galaxy is added to the total extinction, can be studied.
According to the measurements of \citet{men09a} the mean extinction profile around the galaxies in their sample, 
which span the brightness range $17 < m_i < 21$, follow 
\begin{equation}
\langle A_V\rangle (\xi)=4.14\times10^{-3}
\left(\frac{\xi}{100h^{-1} \mbox{ kpc}}\right)^{-0.84}.
\end{equation}
The extinction in the $B$-band can be obtained from the $V$-band extinction via 
$A_B =(R_B/R_V)A_V $. We assume the value $R_V=3.1$ corresponding to Milky Way dust. 
For our sample of SNe~Ia the more detailed model predicts larger extinctions ($\langle A_B \rangle=0.05$ mag) 
than the redshift dependent model ($\langle A_B \rangle=0.03$ mag). 
Also for this model the 
differences in $\Omega_{\rm M}$, $\mathcal{M}$, $\alpha$, and $\beta$ are small compared to the uncorrected case (see table~\ref{tab:fit}).
For the residuals the mean difference is $0.000 \pm 0.009$ mag.
The value of $\beta$ is, however, slightly larger than for the redshift dependent extinction model.
The halo parameters $\sigma_*^{\rm p}$ and  $\sigma_*^{\rm sf}$ are slightly increased by 
10 km s$^{-1}$ each, when the extinction corrected residuals are used in the fitting procedure. 

Consequently, neither of the two dust extinction models have a significant  impact on the results. We conclude therfore that dust extinction is not a plausible explain for the small values of the velocity dispersion normalisations we find for $r_{\rm t}=r_{\rm vir}$.

\begin{table*}
\caption{Differences in cosmological parameters due to dust extinction correction compared to the uncorrected case.} 
\label{tab:fit}      
\centering                                      
\begin{tabular}{l c c r@{.}l c}          
\hline\hline                        
Extinction correction & $\Delta \alpha$& $\Delta \beta$ & \multicolumn{2}{c}{$\Delta \mathcal{M}$} & $\Delta \Omega_{\rm M}$
\\ %
\hline                                   
Redshift dependent & $0.002$ ($0.03\sigma$)& $0.003$ ($0.03\sigma$)& $0$ & $001$ ($0.08\sigma$) & 
$0.009$ ($0.50\sigma$) \\ 
From individual galaxies & $0.003$ ($0.04\sigma$)& $0.026$ ($0.26\sigma$) & $-0$ & $001$ ($0.08\sigma$) & 
$0.007$ ($0.39\sigma$) \\ 
\hline                                             
\end{tabular}
\end{table*}

\section{Gravitational lensing brightness scatter} \label{sec:impl}
The results of the previous sections allow us to investigate the implications for supernova cosmology. 
We restrict ourselves to the model investigated in section~\ref{sec:dis},  where passive and star forming galaxies are described by different velocity dispersion scaling laws. This model is more realistic than the ones where all galaxies are treated as if they were of the same type.

\begin{figure} 
\includegraphics[width=84mm,angle=-90]{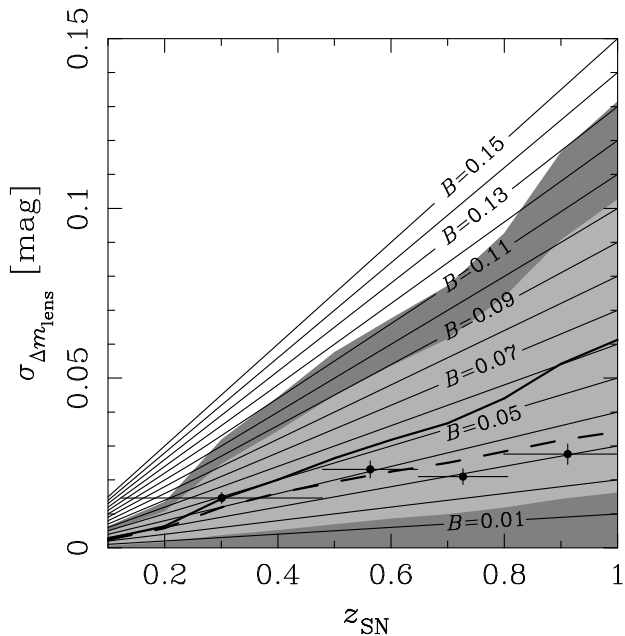}
\caption{
Gravitational lensing scatter as a function of redshift.
The thick solid curve shows the dispersion for the best-fitting model found in section~\ref{sec:dis}. Light and dark shaded regions correspond to 68.3 and 95 per cent confidence level, respectively.
Predictions for a simple model of the lensing scatter, 
$\sigma_{\Delta m_{\rm lens}}=B z_{\rm SN}$, for $B=0.01, 0.02, \ldots,0.15$ mag are indicated by the thin lines. The
circles show the scatter in the magnification computed in four redshift bins for the sample of SNe~Ia used in the analysis. Horizontal error bars indicate the width of the redshift bins.
The thick dashed curve shows the dispersion for the best-fitting model when SNe~Ia with $\Delta m_{\rm SN} < -0.25$ mag have been discarded.
}
\label{fig:glscat} 
\end{figure}

As already stated in the introduction, gravitational magnification leads to additional SN~Ia brightness scatter. This scatter is expected to increase with redshift. The thick solid curve in Fig.~\ref{fig:glscat} indicates the dispersion in $\Delta m_{\rm lens}$
as a function of redshift computed for the best-fitting model after marginalisation over $r_*$
using randomly selected lines of sight. We use the best-fitting truncation radius, $r_*=45h^{-1}$ kpc.
The light and dark shaded regions indicate the scatter at the 68.3 and 95 per cent confidence level translated from Fig.~\ref{fig:fit2}.

Using simulations \citet{hol05} predicted the increase in lensing 
dispersion to be approximately proportional to the SN~Ia  redshift, $\sigma_{\Delta m_{\rm lens}}=B z_{\rm SN}$, with $B=0.088$ mag.  
In Fig.~\ref{fig:glscat} we show the predicted dispersion for constants of proportionality
in the range $B=0.01,0.02,\ldots,0.15$ mag. From the figure we conclude that it is reasonable to describe the redshift dependence of the lensing scatter as proportional to the SN~Ia redshift.
Other authors have found lower values  for the dispersion than \citet{hol05}.
\citet{ber00} find $\sigma_{\Delta m_{\rm lens}} \simeq 0.04$ mag for $z_{\rm SN}=1$  using simulations and assuming
smooth halo profiles. 

From Fig.~\ref{fig:glscat} we find $B \simeq 0.055_{-0.041}^{+ 0.039}$ mag.
At the 95 per cent confidence level we find $B \la 0.12$ mag. 
The results are thus consistent with the simulations of \citet{hol05} and \citet{ber00} at the 68.3 per cent confidence level.

The value of the gravitational lensing dispersion is very sensitive to the high magnification tail. For small samples of SNe~Ia, which fail to sample the high magnification tail, we might therefore find a smaller dispersion. This appears to be the case for our sample of 175 SNe~Ia. The circles in 
 Fig.~\ref{fig:glscat} shows the lensing scatter of these SNe~Ia computed for four redshift bins with $\simeq 44$ objects in each bin. 
The circles are consistent with $B \simeq 0.035$ mag, which is smaller than the value predicted by the best-fitting model. Rejecting simulated SNe~Ia with 
$\Delta m_{\rm lens} < -0.25$ mag leads to the dispersion shown by the thick dashed curve, which agrees with the scatter for the SN~Ia sample.
Only 0.3 (0.7) per cent of the SNe~Ia would be brighter than $-0.25$ mag for SNe with redshift 0.7 (1.0). 
 For the whole sample of 243 SNe~Ia the best-fitting model predicts the number of supernovae brighter than $-0.25$ mag to be 0.8. For the subsample of SNe~Ia used in the lensing analysis the corresponding number is 0.6.
 Since the number of SNe~Ia belonging to the tail is  expected to be very small, our sample of SNe~Ia is probably not sampling the high redshift tail brighter than $\Delta m_{\rm lens} \simeq -0.25$ mag.

\section{Discussion and summary} \label{sec:disc}
We have used presumably  lensed SNe~Ia from 3-year of the SNLS to investigate properties of dark matter haloes of galaxies in the deep CFHTLS  fields.
The dark matter haloes were modeled as truncated singular isothermal spheres with velocity dispersion and truncation radius given by 
luminosity dependent scaling laws. 

Another way to 
probe dark matter haloes is via galaxy--galaxy lensing \citep{tys84,bra96,hud98,fis00,mck01,guz02,hoe03,hoe04,kle06,man06,par07,man08,man09}, which relies upon
measurement of shear via the ellipticity of lensed galaxies rather than the convergence. 
In the galaxy--galaxy lensing literature different truncation radius scaling laws of the form expressed by equation~(\ref{eq:trlaw}) have been explored. 
\citet{hoe03} considered, for example,  the size of galaxy haloes to be 
either the same for all galaxies ($\gamma=0$) or scale as $\sigma^2$ ($\gamma=2$). Since the virial radius of a SIS is proportional to its velocity dispersion, another plausible 
scaling law would be $\gamma=1$.

Fixing the exponent of the velocity dispersion scaling law, equation~(\ref{eq:sislaw}), to $\eta=1/3$ 
we tried to constrain the truncation radius scaling law,
but the data did not allow us to constrain $r_*$ or $\gamma$.

We have also explored the velocity dispersion scaling law by setting $\gamma=0$. Figure~\ref{fig:fit1} shows the result of this exercise. 
\citet{hoe04} used galaxy--galaxy lensing to investigate the properties of dark matter. They used a model slightly different from ours, namely 
 the truncated isothermal sphere model
\citep{bra96}, 
which has a parameter  $s$ describing the truncation scale of the halo. For a value of $\eta$ fixed to 0.3 they found 
$\sigma_*=136 \pm 5 \pm 3$ km s$^{-1}$ (statistical and systematic uncertainties) and $s=185^{+30}_{-28}h^{-1}$ kpc. 
\citet{par07} found a similar value of the velocity dispersion, $\sigma_*=137 \pm 11$ km s$^{-1}$, using CFHTLS data.
In another  galaxy--galaxy lensing study \citet{kle06} found $\sigma_*=132^{+18}_{-24}$ km s$^{-1}$ and $\eta=0.37 \pm 0.15$ 
for SIS haloes truncated at $350h^{-1}$ kpc. Our results for the velocity dispersion scaling law are in good agreement with the findings of these studies. The data prefer, however, a much smaller truncation radius.

Moreover, we tried to disentangle the scaling laws of passive and star forming galaxies. Again we restricted ourselves to models with $\gamma=0$. Passive and star forming galaxies were assumed to obey F--J and T--F relations, respectively.
The best-fitting values of the velocity dispersion and confidence level contours obtained after marginalisation over $r_*$ 
are shown in Fig.~\ref{fig:fit2}.  
The point $\sigma_*^{\rm p}=\sigma_*^{\rm sf}=0$ in this plot, which corresponds to no lensing, is excluded at the 91.6 per cent confidence level.

The results agree well with empirical F--J and T--F relations.
\citet{mit05} measured the F--J relation in the $r$-band using Sloan Digital Sky Survey data. Their measurement corresponds to a velocity dispersion
normalisation of $190$ km s$^{-1}$ for our choice of $L_*$, i.e.~$M_B=-20.3$ mag in the Vega system. 
\citet{boe04} derived a redshift dependent T--F relation using data from \citet{pie92}. For a galaxy at $z=0$ their findings corresponds to a normalisation of 122 km s$^{-1}$.
 
Assuming the truncation radius to be given by the virial radius we found  $\sigma_*^{\rm p}< 195$ and $\sigma_*^{\rm sf}< 117$ km s$^{-1}$, a result which is marginally consistent with haloes truncated at the virial radius with velocity dispersion given by F--J and T--F relations. 
Correcting the SNe~Ia for dust extinction using the results of \citet{men09a} and \citet{men09b} have no significant impact on 
$\sigma_*^{\rm p}$ and $\sigma_*^{\rm sf}$.

Furthermore,  we have used the best-fitting model with different scaling laws for passive and star forming galaxies 
 to investigate the gravitational lensing scatter. For a model where the scatter is proportional to source redshift,
  $\sigma_{\Delta m_{\rm lens}}=B z_{\rm SN}$, we 
 find $B \simeq 0.055_{-0.041}^{+ 0.039}$ mag (at 68.3 per cent confidence level and $B \la 0.12$ mag 
at the 95 per cent confidence level). 
 The scatter is consistent with the prediction of 
\citet[$B=0.088$ mag,][]{hol05} 
at the 68.3 per cent confidence level. 

If the best-fitting model gives an accurate description of the gravitational lensing of  SNe~Ia, the contribution from gravitational lensing scatter to the intrinsic dispersion ($\simeq0.09$ mag) of the SNLS SNe~Ia is very small. 
For SN~Ia data sets with $z_{\rm SN} \ga 1.6$, gravitational lensing could, according to the best-fitting model, contribute significantly to the brightness scatter. The farthest known SN~Ia (1997ff) with  $z_{\rm SN} \simeq 1.8$ is indeed believed to be magnified by a few tenths of a magnitude \citep{lew01,rie01,mor01b,ben02,jon06}.

\section*{acknowledgements}
JJ would like to thank Taia Kronborg, Julien Guy, Pierre Astier, Edvard M\"ortsell, and Ariel Goobar for helpful discussions. 
MS acknowledges support from the Royal Society.

Based on observations obtained with MegaPrime/ MegaCam, a joint project
of CFHT and CEA/DAPNIA, at the Canada-France-Hawaii Telescope (CFHT)
which is operated by the National Research Council (NRC) of Canada,
the Institut National des Science de l'Univers of the Centre National
de la Recherche Scientifique (CNRS) of France, and the University of
Hawaii. This work is based in part on data products produced the
Canadian Astronomy Data Centre as part of the Canada-France-Hawaii
Telescope Legacy Survey, a collaborative project of NRC and CNRS.

Based on observations obtained at the Gemini Observatory, which is operated by the
Association of Universities for Research in Astronomy, Inc., under a cooperative agreement
with the NSF on behalf of the Gemini partnership: the National Science Foundation (United
States), the Science and Technology Facilities Council (United Kingdom), the
National Research Council (Canada), CONICYT (Chile), the Australian Research Council
(Australia), MinistŽrio da Cincia e Tecnologia (Brazil) 
and Ministerio de Ciencia, Tecnolog'a e Innovaci—n Productiva  (Argentina)
%
The programs under which data were obtained at the Gemini Observatory are:
GS-2003B-Q-8,
GN-2003B-Q-9,
GS-2004A-Q-11,
GN-2004A-Q-19,
GS-2004B-Q-31,
GN-2004B-Q-16,
GS-2005A-Q-11,
GN-2005A-Q-11,
GS-2005B-Q-6,
GN-2005B-Q-7,
GN-2006A-Q-7, and
GN-2006B-Q-10.

Based in part on observations made with ESO Telescopes at the Paranal
Observatory under program IDs 171.A-0486 and 176.A-0589.  

Some of the
data presented herein were obtained at the W.M. Keck Observatory,
which is operated as a scientific partnership among the California
Institute of Technology, the University of California and the National
Aeronautics and Space Administration. The Observatory was made
possible by the generous financial support of the W.M. Keck
Foundation.



\begin{thebibliography}{199}
%
\bibitem[\protect\citeauthoryear{Astier et al.}{2006}]{ast06}
Astier  P., Guy J., Regnault N., et al., 2006, A\&A, 447, 31

\bibitem[\protect\citeauthoryear{Balland et al.}{2009}]{bal09}
Balland C., Baumont S., Basa S., et al., 2009, A\&A, 507, 85

\bibitem[\protect\citeauthoryear{Ben\'itez et al.}{2002}]{ben02}
Ben\'itez N., Riess A., Nugent P., Dickinson M., Chornock R., Filippenko A.,
2002, ApJ, 577, L1

\bibitem[\protect\citeauthoryear{Bergstr\"om et al.}{2000}]{ber00}
Bergstr\"om L., Goliath  M., Goobar  A.,  M\"ortsell  E., 2000, A\&A, 358, 13

\bibitem[\protect\citeauthoryear{Bertin \& Arnouts}{1996}]{ber96}
Bertin E., Arnouts S., 1996, A\&AS 117, 393

\bibitem[\protect\citeauthoryear{B\"ohm et al.}{2004}]{boe04}
B\"ohm A., Ziegler B.L., Saglia R.P, et al., 2004, A\&A, 420, 97

\bibitem[\protect\citeauthoryear{Boulade et al.}{2003}]{bou03}
Boulade O., Charlot X., Abbon P., et al., 2003, in Instrument Design and Performance 
for Optical/Infrared Ground-based Telescopes, ed. Iye, Masanori; Moorwood, Alan F.~M., Proc. SPIE, 4841, 72

\bibitem[\protect\citeauthoryear{Brainerd, Blandford \& Smail}{1996}]{bra96}
Brainerd T.G., Blandford R.D.,  Smail I. ,1996, ApJ, 466, 623

\bibitem[\protect\citeauthoryear{Bronder et al.}{2008}]{bro08}	
Bronder T.J., Hook I.M., Astier P., et al.,  2008, A\&A, 477, 717

\bibitem[\protect\citeauthoryear{Bryan \& Norman}{1998}]{bry98}
Bryan G., Norman M., 1998, ApJ, 495, 80


\bibitem[\protect\citeauthoryear{Cardelli, Clayton \& Mathis}{1989}]{car89}
Cardelli J.A., Clayton G.C., Mathis J.S., 1989, ApJ, 345, 245


\bibitem[\protect\citeauthoryear{Conley et al.}{2008}]{con08}
Conley A., Sullivan M., Hsiao E.Y., et al., 2008, ApJ, 681, 482

\bibitem[\protect\citeauthoryear{Dodelson \& Vallinotto}{2006}]{dod06}
Dodelson S., Vallinotto A.,  2006, Phys. Rev. D, 74, 063515

\bibitem[\protect\citeauthoryear{Faber \& Jackson}{1976}]{fab76}
Faber S.M., Jackson R.E., 1976, ApJ, 204, 668

\bibitem[\protect\citeauthoryear{Fischer et al.}{2000}]{fis00}
Fischer P., et al., 2000, AJ, 120, 1198

\bibitem[\protect\citeauthoryear{Fioc \& Rocca-Volmerange}{1997}]{fioc97}
Fioc M., Rocca-Volmerange B., 1997, A\&A 326, 950

\bibitem[\protect\citeauthoryear{Frieman}{1997}]{fri97}
Frieman J., 1997, Comments Astrophys., 18, 323

\bibitem[\protect\citeauthoryear{Gunnarsson et al.}{2006}]{gun06}
Gunnarsson C., Dahl\'en T., Goobar A., J\"onsson J., M\"ortsell E., 2006, ApJ, 640, 471

\bibitem[\protect\citeauthoryear{Guy et al.}{2007}]{guy07}
Guy J., Astier P., Baumont S., et al., 2007, A\&A, 466, 11

\bibitem[\protect\citeauthoryear{Guy et al.}{2009}]{guy09}
Guy J., et al., 2009, in preparation

\bibitem[\protect\citeauthoryear{Guzik \& Seljak}{2002}]{guz02}
Guzik J., Seljak U., 2002, MNRAS, 355, 311

\bibitem[\protect\citeauthoryear{Hoekstra et al.}{2003}]{hoe03}
Hoekstra H., Franx M., Kuijken K., Carlberg R.G., Yee H.K.C., 2003, MNRAS, 340, 609

\bibitem[\protect\citeauthoryear{Hoekstra et al.}{2004}]{hoe04}
Hoekstra H., Yee H.K.C., Gladders M.D., 2004, ApJ, 606, 67

\bibitem[\protect\citeauthoryear{Holz \& Linder}{2005}]{hol05}
Holz D.E., Linder E.V., 2005, ApJ, 631, 678

\bibitem[\protect\citeauthoryear{Holz \& Wald}{1998}]{hol98}
Holz D.E., Wald R.M., 1998, Phys. Rev. D, 58, 063501

\bibitem[\protect\citeauthoryear{Howell et al.}{2005}]{how05}
Howell D.A., Sullivan M., Perrett K., et al, 2005, ApJ, 634, 1190

\bibitem[\protect\citeauthoryear{Hudson et al.}{1998}]{hud98}	
Hudson M.J., Gwyn S.D.J., Dahle H., Kaiser N., 1998, ApJ, 503, 531

\bibitem[\protect\citeauthoryear{Ilbert et al.}{2006}]{ilb06}
Ilbert O., Arnouts S., McCracken H.J., et al., 2006, A\&A, 457, 841

\bibitem[\protect\citeauthoryear{J\"onsson et al.}{2006}]{jon06}
J\"onsson J., Dahl\'en T., Goobar A., Gunnarsson C., M\"ortsell E.,  Lee K., 
2006, ApJ, 639, 991

\bibitem[\protect\citeauthoryear{J\"onsson et al.}{2007}]{jon07}
J\"onsson J., Dahl\'en T., Goobar A., M\"ortsell E.,  Riess A., 2007, 
JCAP, 06, 002

\bibitem[\protect\citeauthoryear{J\"onsson et al.}{2008}]{jon08}
J\"onsson J., Kronborg T., M\"ortsell E., Sollerman J., 2008, 
A\&A, 487, 467

\bibitem[\protect\citeauthoryear{J\"onsson et al.}{2009}]{jon09}
J\"onsson J., Dahl\'en T., Hook I., Goobar A., M\"ortsell E.,  2009, preprint (arXiv:0910.4098)

\bibitem[\protect\citeauthoryear{J\"onsson, M\"ortsell \& Sollerman}{2009}]{jon09b}
J\"onsson J., M\"ortsell E., Sollerman J., 2009, 
A\&A, 493, 331

\bibitem[\protect\citeauthoryear{Kantowski, Vaughan \& Branch}{1995}]{kan95}
Kantowski R., Vaughan T., Branch D., 1995, ApJ, 447, 35

\bibitem[\protect\citeauthoryear{Kleinheinrich et al.}{2006}]{kle06}
Kleinheinrich M., et al., 2006, A\&A, 455, 441

\bibitem[\protect\citeauthoryear{Kronborg et al.}{2010}]{kro09}
Kronborg T., et al., 2010, submitted

\bibitem[\protect\citeauthoryear{Le Borgne \& Rocca-Volmerange}{2002}]{leborgne02}
Le Borgne D., Rocca-Volmerange B., 2002, A\&A 386, 446

\bibitem[\protect\citeauthoryear{Lewis \& Ibata}{2002}]{lew01}	
Lewis G., Ibata R., 2001, MNRAS, 324, L25

\bibitem[\protect\citeauthoryear{Mandelbaum et al.}{2006}]{man06}	
Mandelbaum R., Seljak U., Kauffmann G., Hirata C.M., Brinkmann J., 2006, MNRAS, 368, 715

\bibitem[\protect\citeauthoryear{Mandelbaum, Uros \& Hirata}{2008}]{man08}
Mandelbaum R., Seljak U., Hirata C.M. 2008, JCAP, 08, 006

\bibitem[\protect\citeauthoryear{Mandelbaum et al.}{2009}]{man09}
Mandelbaum R., Li C., Kauffmann G., White S.D.M., 2009, MNRAS, 393, 377

\bibitem[\protect\citeauthoryear{McKay et al.}{2001}]{mck01}
McKay T.A., Sheldon E.S., Racusin, J., et al., 2001, preprint (astro-ph/0108013)

\bibitem[\protect\citeauthoryear{M\'enard et al.}{2009}]{men09a}
M\'enard B., Scranton R., Fukugita M., Richards G., 2009, preprint 
(arXiv:0902.4240) 

\bibitem[\protect\citeauthoryear{M\'enard, Kilbinger \& Scranton}{2009}]{men09b}
M\'enard B., Kilbinger M., Scranton R., 2009, preprint 
(arXiv:0903.4199) 

\bibitem[\protect\citeauthoryear{Metcalf}{1998}]{met98}
Metcalf R.B., 1998, in Banday A., Sheth R., da Costa L., eds, Evolution of 
Large-Scale Structure: From Recombination to Garching, preprint 
(astro-ph/9810440) 


\bibitem[\protect\citeauthoryear{Metcalf}{1999}]{met99}
Metcalf R.B., 1999, MNRAS, 305, 746

\bibitem[\protect\citeauthoryear{Metcalf \& Silk}{1999}]{metsil99}
Metcalf R.B., Silk J., 1999, ApJ, 519, L1

\bibitem[\protect\citeauthoryear{Metcalf \& Silk}{2007}]{met07}
Metcalf R.B., Silk J., 2007, Phys. Rev. Lett., 98, 071302

\bibitem[\protect\citeauthoryear{Minty et al.}{2002}]{min02}
Minty  E.M., Heavens A.F., Hawkins M.R.S., 2002, MNRAS, 330, 378

\bibitem[\protect\citeauthoryear{Mitchell et al.}{2005}]{mit05}
Mitchell J., Keeton C., Frieman  J., Sheth R., 2005, ApJ, 622, 81

\bibitem[\protect\citeauthoryear{M\"ortsell, Goobar \& Bergstr\"om}{2001}]{mor01}
M\"ortsell E., Goobar A., Bergstr\"om L., 2001, ApJ, 559, 53

\bibitem[\protect\citeauthoryear{M\"ortsell, Gunnarsson \& Goobar}{2001}]{mor01b}
M\"ortsell E., Gunnarsson C., Goobar A., 2001, ApJ, 561, 106

\bibitem[\protect\citeauthoryear{Parker et al.}{2007}]{par07}	
Parker L.C., Hoekstra H., Hudson M.J., van Waerbeke L., Mellier Y.,
2007, ApJ, 669, 21

\bibitem[\protect\citeauthoryear{Pierce \& Tully}{1992}]{pie92}
Pierce M.J., Tully R.B., 1992, ApJ, 387, 47

\bibitem[\protect\citeauthoryear{Rauch}{1991}]{rau91}
Rauch K.P., 1991, ApJ, 374, 83

\bibitem[\protect\citeauthoryear{Rieke \& Lebofsky}{1985}]{rie85}
Rieke G.H., Lebofsky M.J., 1985, ApJ, 288, 618

\bibitem[\protect\citeauthoryear{Riess et al.}{2001}]{rie01}
Riess A., et al., 2001, ApJ, 560, 49

\bibitem[\protect\citeauthoryear{Sarkar et al.}{2008}]{sar08}
Sarkar D., Amblard A., Holz D.E., Cooray A.,
2008, ApJ, 678, 1

\bibitem[\protect\citeauthoryear{Savage \& Mathis}{1979}]{sav79}
Savage B.D., Mathis J.S., 1979, Ann. Rev. Astr. Ap., 17, 73

\bibitem[\protect\citeauthoryear{Schneider, Ehlers \& Falco}{1992}]{sch92}
Schneider P., Ehlers J., Falco E.E., 1992, Gravitational Lenses, Springer-Verlag, Berlin 

\bibitem[\protect\citeauthoryear{Scranton et al.}{2005}]{scr05}	
Scranton R., M\'enard B.,  Richards G.T., et al., 2005, ApJ, 633, 589

\bibitem[\protect\citeauthoryear{Seaton}{1979}]{sea79}
Seaton M.J., 1979, MNRAS, 187, 73

\bibitem[\protect\citeauthoryear{Seljak \& Holz}{1999}]{sel99}
Seljak U., Holz D.E., 1999, A\&A, 351, L10

\bibitem[\protect\citeauthoryear{Sullivan et al.}{2006}]{sul06}
Sullivan M., et al., 2006, ApJ 648, 868

\bibitem[\protect\citeauthoryear{Tripp}{1998}]{tri98}
Tripp R., 1998, A\&A, 331, 815

\bibitem[\protect\citeauthoryear{Tully \& Fisher}{1977}]{tul77}
Tully R.B., Fisher J.R., 1977, A\&A, 54, 661

\bibitem[\protect\citeauthoryear{Tyson et al.}{1984}]{tys84}	
Tyson J.A., Valdes F., Jarvis J.F., Mills A.P. Jr, 1984, ApJ, 281, L59

\bibitem[\protect\citeauthoryear{Wambsganss et al.}{1997}]{wam97}
Wambsganss J., Cen R., Xu G.,  Ostriker J.P.,
1997, ApJ, 475, L81

\bibitem[\protect\citeauthoryear{Zentner \& Bhattacharya}{2009}]{zen09}
Zentner A.R., Bhattacharya S., 2009, ApJ, 693, 1543

\end{thebibliography}
\end{document}